\documentclass[osajnl,twocolumn,showpacs,superscriptaddress,10pt]{revtex4-1} 
\usepackage{amssymb}
\usepackage{amsmath}
\usepackage{graphicx}
\usepackage{bm}

\newcommand{\rmd}{\mathrm{d}}
\newcommand{\rme}{\mathrm{e}}
\newcommand{\be}{\begin{equation}}
\newcommand{\ee}{\end{equation}}

\newcommand{\bc}[1]{\bm{\mathcal{#1}}}
\newcommand{\bd}[1]{\mathbf{#1}}

\newcommand{\bs}{\begin{subequations}}
\newcommand{\es}{\end{subequations}}

\newcommand{\half}{{\textstyle\frac1{2}}}
\newcommand{\third}{{\textstyle\frac1{3}}}

\newcommand{\bn}{\bar{n}}
\newcommand{\bna}{\bn\alpha}
\newcommand{\bnar}{\bn\alpha\varrho}
\newcommand{\tc}{\tilde{c}}
\newcommand{\td}{\tilde{d}}
\newcommand{\tA}{\tilde{A}}
\newcommand{\tB}{\tilde{B}}

\newcommand{\re}{\mathrm{Re}}

\begin{document}

\title{Theory of microdroplet and microbubble deformation by Gaussian laser beam}

\author{Simen \AA.\ Ellingsen}\email{simen.a.ellingsen@ntnu.no}
\affiliation{Department of Energy and Process Engineering, Norwegian University of Science and Technology, N-7491 Trondheim, Norway}

\begin{abstract}
  The theory for linear deformations of fluid microparticles in a laser beam of Gaussian profile is presented, when the beam focus is at the particle center as in optical trapping. Three different fluid systems are considered: water microdroplet in air, air microbubble in water, and a special oil-emulsion in water system used in experiments with optical deformation of fluid interfaces. We compare interface deformations of the three systems when illuminated by a wide (compared to particle radius) and narrow laser beams and analyse differences. Deformations of droplets are radically different from bubbles under otherwise identical conditions, due to the opposite lensing effect (converging and diverging, respectively) of the two; a droplet is deformed far more than a bubble, \emph{cetera paribus}. Optical contrast is found to be of great importance to the shape obtained when comparing the relatively low-contrast oil-emulsion system to that of water droplets. We finally analyse the dynamics of particle motion when the laser beam is turned on, and compare a static beam to the case of a short pulse. The very different surface tension coefficient implies a very different time scale for dynamics: microseconds for the water-air interface and tens of milliseconds for the oil-emulsion. Surface oscillations of a water microdroplet are found always to be underdamped, while those of the oil-emulsion are overdamped; deformations of a microbubble can be either, depending on physical parameters.
\end{abstract}

\maketitle 

\section{Introduction}

Optofluidics, the marriage of optics and microfluidics, is a field which has experienced rapid growth in recent years. Many areas of application have already been demonstrated, including chemo-biological applications\cite{guck00,guck01,fan11}, solar energy applications \cite{erickson11}, microdroplet lasing \cite{li08}, optically controlled droplet transport and coalescence \cite{baroud07} which can be a vehicle for DNA calculations \cite{nishimura12} and a considerable range of dynamically configurable integrated fluid-based optophotonic devices \cite{psaltis06,monat07a, monat08}. Lab-on-a-chip applications of microfluidics have already been realised for various purposes \cite{baroud07,schaap11},  For an in-context review, see Ref.~\cite{zappe10}.

Applications of optical trapping and tweezing of microscopic fluid droplets are easy to imagine. Optical tweezers allow detailed manipulation of microsystems \cite{grier03}. A laser can for example measure the size of a spherical droplet via its Mie resonances (an old idea, see e.g.\ Ref.~\cite{sinclair49}) and automatically choose and transport the desired quantities of reactants for microchemistry on a chip. A related technique was alredy demonstrated for sorting of particles by size or refractive index \cite{macdonald03}. 


In light of the technological promise of laser manipulation of microfluidics, it is of obvious interest to consider the deformation and motion of a microdroplet moved by a laser. The classic degrees of freedom when manipulating microobjects with light are pushing and trapping, and recently it was discovered that special laser beams can also pull against the direction of propagation \cite{chen11}. While the fluid droplet behaviour under pushing and pulling have been analysed in the past \cite{ellingsen12}, the present paper has particular relevance to the case of optically trapped droplets, bubbles and emulsions and, even more so, deliberate deformation of droplets, bubbles and emulsion by a Gaussian beam. 

As well as using lasers to move fluids around, controlled deformation of fluids using lasers has attracted considerable interest. 
A number of impressive experiments in recent times demonstrate the potential for using laser light for detailed manipulation of microflows. In particular, two-fluid systems of oil emulsion droplets have been used, in which the surface tension of the liquid--liquid interfaces can be reduced to a millionth of that of water--air. References include a series of works by Delville's group\cite{casner03,schroll07, chraibi10,wunenburger11, issenmann11} and associated theory \cite{hallanger05,birkeland08}, as well as ``optical sculpting'' work at the Central Laser Facility~\cite{ward06,ward07,woods11}. Mitani and Sakai used a similar system for the measurement of surface tension~\cite{mitani02}. 

The manipulation of liquid surfaces with laser light goes back to Ashkin an Dziedzic \cite{ashkin73}. It was soon confirmed \cite{lai76} that a correct description of the observed deformation could be given by integrating the Minkowski electromagnetic stress tensor across the surface.

A classical experiment on the deformation of droplets by laser pulses was that of Zhang and Chang \cite{zhang88}, whose results could be satisfactorily analysed with essentially the same means \cite{lai89, brevik99}, drawing on Mie's theory for light scattering on spherical particles \cite{mie08}. 

The deformation of a liquid droplet with ultra-low surface tension trapped in a laser beam was studied experimentally by M\o ller and Oddershede \cite{moller09} As the droplet is elongated in the direction of the laser propagation, its radius in the cross--beam plane decreases. The radius was measured as a function of laser power for three different droplet sizes and compared to a simple model. A linear decrease with laser power, predicted by their model as well as the one presented herein, was observed for low laser powers, but beyond a threshold power the variation with power becomes much weaker. While still limited to small deformations, the theory presented in this paper is able to elucidate the observations by M\o ller and Oddershede. In particular, their experiment gives an indication of the range of validity of the linear deformation theory.

In the present manuscript we also study the ``opposite'' system exemplified by air bubbles in water, such as considered experimentally in \cite{anand11}. The theory for deformation of bubbles is virtually identical to that for droplets, but the deformation is qualitatively different. In general the case denoted ``bubble'' includes all cases where the illuminated spherical fluid object has a lower refractive index than the surrounding medium, as is the case for air bubbles in water. In section \ref{sec_theory} where we lay out the theoretical framework, we shall only speak of ``droplets'' for simplicity, in the understanding that the exact same expressions are applicable to the bubble situation.

\begin{figure}[htb]
  \includegraphics[width=3in]{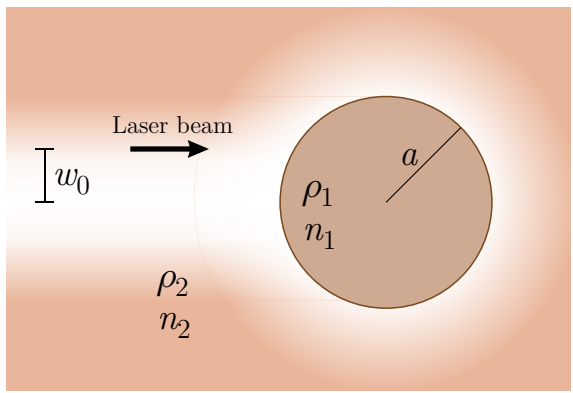}
  \caption{The set-up considered. An initially spherical particle of fluid $1$ enclosed within a fluid $2$ is illuminated by a Gaussian beam so that the particle's center is at the beam focus.}
  \label{fig_geom}
\end{figure}

We consider the set-up shown in Fig.~\ref{fig_geom}. A droplet (bubble) of radius $a$ when unperturbed, density $\rho_2$ and refractive index $n_2$ sits in a medium of density $\rho_1$ and refractive index $n_1$ and is illuminated by a laser beam whose focal waist is $w_0$ which can be bigger or smaller than the droplet radius. The droplet is deformed by the optical force acting upon it. We consider both the case where the laser is pulsed, and, with certain approximations also the case where it is trapped in a static beam.

We work in the analytical framework made possible by Mie's scattering theory \cite{mie08}. Analytical results for the electric field near a deformed sphere do not exist at present, although certain considerations have been made, e.g., for the change in lasing properties \cite{lai91,mekis95}. We will work with the internal and scattered electric laser field as for a spherical droplet whereas all other droplet dynamics take account of deformations to linear order. For the case of a short pulse, this is virtually exact since hydrodynamic motion takes place after the duration of the pulse, while for the case of a trapped droplet it is an approximation, the best such presently available short of direct numerical simulation for which it is harder to keep track of the individual physical effects. In the case of oil-emulsion systems the dielectric contrast is small and the difference in internal diffraction pattern will be modest due to a small deformation, whereas for higher contrast (e.g.\ water in air) a fully extended theory is certainly desirable and we plan to return to this in future work.

The theoretical framework of both the fluid mechanics and the optics of the problem at hand are laid out in Section \ref{sec_theory}. A number of numerical examples are shown and discussed in section \ref{sec_numerics} whereupon concluding remarks are made. 
The effects of gravity (including buoyancy) are neglected throughout the manuscript.

\section{Theoretical framework}\label{sec_theory}

We consider the case of an incident field $\bc{E}^i$ impinging on a droplet with
radius $a$ and
refractive index $n_1$, embedded in a medium of refractive index $n_2$. In the present paper we shall assume $n_1$ and $n_2$ to be real for simplicity, although the generalisation to a droplet of complex $n_1$, hence absorbing droplet, is straightforward. 
We assume all fluids to be nonmagnetic, i.e., having vacuum permeability.

\subsection{Electromagnetic force density}

In general, the electromagnetic (EM) force density acting in an isotropic dielectric medium may be written \cite{brevik79}
\begin{align}
  {\bf f}=&-\frac{1}{2}\epsilon_0\langle\bc{E}^2\rangle\nabla n^2 +\frac{1}{2}\nabla\left[ \langle\bc{E}^2\rangle\rho \left(\frac{\partial n^2}{\partial \rho}\right)_T\right]\notag \\
  &+\frac{n^2-1}{c^2}\frac{\partial}{\partial t}(\bc{E}\times \bc{H}) \label{fE}.
\end{align}
A discussion of the three terms of Eq.~(\ref{fE}) for the present context is found in Ref.~\cite{lai89}.
The last term of Eq.~(\ref{fE}) is called the Abraham term. Its interpretation has been debated for a century \cite{brevik79}, but this is of no concern in the present investigation; since the time variations of our EM fields are at optical frequencies, immensely quicker than the mechanical response time of the fluids, the Abraham term averages to zero over an optical period and is of no consequence, hence we omit it in the following.

The second term of Eq.~(\ref{fE}) is called the electrostrictive term \cite{brevik79}. It may be interpreted as a tendency for the material medium to be attracted to areas of higher field intensity. When the EM field is switched on abruptly, the fluid will be slightly compressed in areas of higher intensity and vice versa. Soon the material itself will counter this compression process with an elastic counter-pressure until mechanical equilibrium is reached on a timescale of the time a sound wave takes to traverse the fluid. Once equilibrium is reached, the electrostriction term is exactly cancelled by a corresponding rise in hydromechanical pressure and is absorbed therein (this point was discussed in the present context in Ref.~\cite{lai89}). What remains is a very slight non-uniformity of the fluid density due to the pressure distribution giving rise, in principle, to a slightly nonlinear optical response which we shall neglect. Ignoring the strictive term as we shall in the following thus means the theory is valid only for timescales longer than the traversal time $2a/c_s$, $c_s$ being the speed of sound. For water and air (typefying a liquid and a gas) this means that our timescale must is restricted to being greater than 
\begin{align*}
  t \gtrsim& 1.4 \mathrm{ns} \frac{a}{1\mu\mathrm{m}}, ~~\text{liquid droplet} \\
  t \gtrsim& 5.9 \mathrm{ns} \frac{a}{1\mu\mathrm{m}}, ~~\text{gas bubble}.
\end{align*}
When considering very short pulses this should be born in mind (for further discussion see Ref.¨\cite{ellingsen12b}). For the present endeavour we shall assume all pulses to be much longer in duration than these timescales.

A final note on the EM force density concerns a suggestion by Peierls \cite{peierls76} that the gradient term (first term of Eq.~(\ref{fE})) should contain a quadratic term in the refractive index, $\propto (\epsilon-1)^2$.
The nonlinear term was expressly omitted in \cite{lai76}, where its effect was estimated to 10\% for water. What to make of this has been a matter of debate, however, as it was found to be inconsistent with an experiment by Jones and Leslie \cite{jones78}, as also detailed in the appendix of Ref.~\cite{brevik79}. A force density nonlinear in refractive index signifies a nonlinear optical effect, hinting that the term has to do with the material's compressibility. Indeed, the apparent mismatch involved would appear to have been resolved by Lai et al.\ \cite{lai81} who show that, just as for electrostriction, the nonlinear force term is exactly cancelled by an increase in pressure, as long as the latter has time to build. It is a good example of how subtle it can be to go from microscopic to macroscopic description in systems where electromagnetic and mechanical forces act together. 

We shall therefore consider only the first term of the force density \eqref{fE}, often referred to as the gradient force, which gives a contribution only at the fluid surface, where $\varepsilon$ varies. It is equivalent (e.g.\ \cite{landau84} \S 59) to
%
integrating the gradient of the Maxwell stress tensor
\be
  \bd{T}=\bc{E}\otimes\bc{D}+\bc{H}\otimes\bc{B}-\half(\bc{E}\cdot\bc{D}+\bc{H}\cdot\bc{B})\bd{1}
\ee
across the surface,
\be
  \sigma(\Omega)=\langle \sigma_{rr}\rangle = \langle T_{rr}(r=a^+)-T_{rr}(r=a^-)\rangle.
\ee
Here $\bd{1}$ is the unit matrix and $\langle \cdots\rangle$ denotes time average over an optical period. 
Since $\mu=1$ everywhere, the magnetic terms of the Maxwell tensor do not contribute to the surface force and may be omitted.

We introduce complex fields so that $\bc{E} = \re\{\bd{E} \rme^{i\omega t}\}$, etc., where $\bd{E},\bd{D},\bd{B}$ and $\bd{H}$ are complex field vectors (denoted as upright, as opposed to calligraphic typeface). For field components $\mathcal{X}_i$ and $\mathcal{X}_j$, we have $\langle \mathcal{X}_i\mathcal{X}_j\rangle = \half \re\{
X_iX_j^*\}$, and in particular $\langle \mathcal{X}_i^2\rangle = \half |X_i|^2$.

The force density on a sphere is found from Maxwell's stress tensor. In terms of the (complex) EM fields just inside the spherical surface it may be writen \cite{ellingsen12}
\be\label{sigmaEM}
   \sigma(\theta)=\frac{\varepsilon_0n_2^2}{4} (\bn^2-1)(\bn^2|E^w_r|^2+|E^w_\theta|^2+|E^w_\phi|^2).  
\ee
where $\bn = n_1/n_2$. Superscript $w$ indicates that the field is evaluated inside the droplet. We shall be working with circularly polarised light so that our system is axially symmetric throughout. Projecting onto a basis of spherical harmonics, we may write in this case,
\be\label{sigmaLeg}
  \sigma(\theta)=\sum_{l=0}^\infty \sigma_l P_l(\cos\theta).
\ee
It was found in Ref.~\cite{ellingsen12} that the $l=0$ term is zero due to mass conservation and that the $l=1$ term corresponds to uniform movement of the entire droplet and can be ignored, so only $l\geq 2$ are of interest to us. The orthogonality relation for Legendre polynomials gives~\cite{ellingsen12}
\be\label{sigmal}
  \sigma_l = \half(2l+1)\int_0^\pi \rmd\theta\sin\theta\sigma(\theta)P_l(\cos\theta).
\ee

We define here for future reference the quantity $\alpha$ which is the number of wavelengths (in the external medium) per circumference, a key parameter,
\be
  \alpha = k_2a = \frac{2\pi a}{\lambda_2} = \frac{n_2\omega a}{c}
\ee
($k_2$: wave number in outer medium).

In this paper we have assumed all fluid to have zero absorption of light for simplicity. Before going on to the fluid mechanics of the droplet, let us briefly discuss the ramifications of this assumption. Two notable physical effects would manifest themselves were we to include a small absorption coefficient for the fluids (equivalent to letting refractive index $n$ have a small positive imaginary part). First, the net force on the droplet would acquire an addition in the direction of propagation due to absorption of photons, each carrying a momentum $\hbar \omega$. More interestingly for us, however, is the photoacoustic effect resulting from the increase in temperature which would result from absorbed energy, causing local thermal expansion of the fluid. The effect has many applications in metrology \cite{tam86}. When the laser is switched on the local temperature and pressure will rise proportional to time $t$ and absorption coefficient $\alpha_\text{abs}$. After a while, local thermal equilibrium is reached -- heat is transported away at the same rate as it is produced -- and the temperature rise and expansion stops. Typical temperature rise and fall times for micrometer sized liquid systems are about $5-10\mu$s \cite{peterman03, cordero09}. The pressure change creates an acoustical wave which propagates out of the system. After the rise time, thermal expansion, just like electrostriction, will contribute only to the pressure distribution, and not to the motion of the fluid. It can thus have a bearing on the fluid surface dynamics only when light is modulated on a timescale similar to the rise time, or short enough for acoustics to play a role.

\subsection{Fluid mechanics}

The optical force acting on the interface between the two immiscible fluids will deform the surface. The deformation can be described transiently from the laser is switched on~\cite{lai89,brevik99,ellingsen12}, but in the present case we shall assume static conditions. We assume throughout that the surface deformation is small enough so that we may keep only linear order in the deformation amplitude. On the other hand we will calculate the optical force as if the droplet/bubble were spherical, which is a reasonable approximation when the dielectric contrast is low, $n_2-n_1\ll n_1$, but corrections of linear order in amplitude are expected to be of greater importance for greater contrast, such as for a water droplet in air as considered in Refs.~\cite{lai89,brevik99,ellingsen12}. (Note that these references consider the case of an optical pulse short enough so that the particle remains spherical throughout its duration). A full theory for the static deformation of trapped microdroplets and -bubbles is under development.

We write the surface deformation of the droplet also in terms of Legendre polynomials:
\be
  r(\theta,t) = a+\sum_{l=2}^\infty h_l(t) P_l(\cos\theta).
\ee
In Ref.~\cite{ellingsen12} was considered the transient case $h_l(t)$ where a laser beam was switched on at $t=0$ and off again after a time $t_0$. The deformation coefficients in the current, static case is then given from the solution found in Eq.~(10) of \cite{ellingsen12} upon taking $t\to \infty$ while keeping $t<t_0$ (a long time after switch--on, but before switch-off). The static solution is 
\be\label{hl}
  h_l(\infty) = \frac{\sigma_l a^2}{\gamma}\frac1{l^2+l-2}
\ee
where $\gamma$ is the surface tension coefficient.

More generally, a time $t>0$ after the beam is switched on, the surface deformation is described for the underdamped and overdamped case, respectively, by \cite{ellingsen12}
\bs\label{motion}
  \begin{align}
   \frac{h_{l}(t)}{h_l(\infty)} =& 1-\Bigl(\frac{\mu_l}{\gamma_l}\sin\gamma_lt+\cos\gamma_lt\Bigr)\rme^{-\mu_l t}, ~~ \omega_l>\mu_l \\
    \frac{h_{l}(t)}{h_l(\infty)} =& 1-\Bigl(\frac{\mu_l}{\Gamma_l}\sinh\Gamma_lt+\cosh\Gamma_lt\Bigr)\rme^{-\mu_l t}, ~~ \omega_l<\mu_l 
  \end{align}
\es
for $ l\geq 2$, where $\Delta \rho=|\rho_2-\rho_1|$ is the difference in mass density and
\bs
  \begin{align}
    \gamma_l =& \sqrt{\omega_l^2-\mu_l^2};~~ \Gamma_l = \sqrt{\mu_l^2-\omega_l^2}\\
    \omega_l^2=&\frac{\gamma l}{\Delta\rho a^3}(l^2+l-2)\\
    \mu_l =& \frac{\nu_1}{a^2}(2l^2-l-1)\frac{1+\frac{\mu_2}{\mu_1}\frac{(l+1)(l+2)}{l(l-1)}}{1+\frac{\rho_2}{\rho_1}\frac{l}{l+1}}.\label{mul}
  \end{align}
\es
The expression for $\mu_l$ generalizes that given in \cite{brevik99,ellingsen12}, which was valid only when the viscosity and density of the outer medium can be neglected compared to the inner. Details of its derivation are found in Appendix \ref{app_Visc}.

The fluid dynamic theory presented is valid to linear order in $h(\theta)$ and $h'(\theta)$. The most important correction to this is likely to be the onset of nonlinear terms from surface tension, since this terms contains linearlised factors $\sqrt{1+[h'(\theta)/a]^2}$, giving third order terms $\sim h[h']^2/a^3$. The quadratic term ignored upon linearising the Navier-Stokes equation for fluid motion, due to internal convection in the droplet, is of gradient form, and may be absorbed in the pressure, being fully accounted for \cite{lai89}. In the plots of fluid shape in the sections to come, the maximum deformations are approximately $|h|/a, |h'|/a \sim 0.2-0.5$, which is pushing the boundaries of the linear regime a little. What is accomplished is a clear presentation of the qualitative shape,  which is the intention of the figures in question, although the actual shapes seen in the figure should be understood as semi-quantitative. Using amplitudes exceeding the linear regime for illustration purposes is standard in the literature, c.f.\ e.g.\ \cite{lai89,brevik99}.

We shall treat both droplets and bubbles as incompressible in the following. For the latter case, this might require comment. For steady flow the condition that compressibility effects be negligible is that the fluid velocity is much smaller than the speed of sound (about $340$m/s in dry air). For the present case, fluid velocities are of order $v\sim a \omega \sim \sqrt{\gamma/\Delta \rho a}$ for underdamped case when velocity is the greatest. Inserting numbers for air surrounded by water, $l=2$, velocities in the order $1$m/s are obtained, hence a Mach number of order $10^{-3}$. For unsteady (transient) flow such as here, a second criterion must be fulfilled (c.f.\ \cite{landau59} \S 10), that the timescale $\tau$ of the motion must be far greater than $L/c_s$, $L$ being a typical lengthscale and $c-s$ the speed of sound. Here, it would imply $a \omega/c_z\ll 1$, again satisfied by two to three orders of magnitude for a micron sized droplet. 

Alternatively, we may estimate the maximum possible change in density by regarding the maximum optical force density acting on the surface, which from Eq.~\eqref{sigmaEM} is $\approx n_2(\bn^2-1)I_0/2c\sim \delta p$, the change in internal pressure. Using the ideal gas law this would correspond to a change in density of $\delta \rho= \delta p/R_\text{air}T$ where the specific gas constant for air is $R_\text{air}=287$J\, kg$^{-1}$\,K$^{-1}$, and $T$ is temperature. With values used for the bubble case we obtain $\delta \rho_\text{max} \approx -0.07$kg/m$^3$ which is tiny compared to atmospheric air density of around $1.2$kg/m$^3$. 
Hence we may safely ignore effects of compressibility for gas-filled bubbles, as well as droplets.

\subsection{Overdamped or underdamped oscillations?}\label{sec_damping}

Oscillations are underdamped whenever $\omega_l^2>\gamma_l^2$ and \emph{vice versa}. We consider three different cases separately: droplet in air, air bubble in liquid, and a liquid-liquid system such as the oil emulsion in water.

\subsubsection{Liquid droplet in gas}

\begin{figure}[tb]
  \includegraphics[width=\columnwidth]{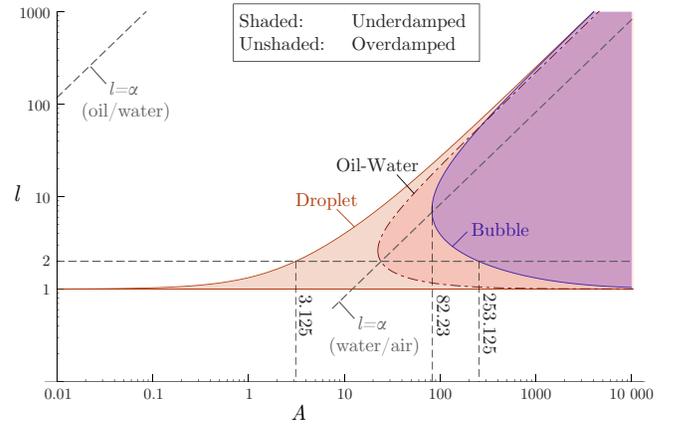}
  \caption{Critical damping curves for the three cases: (1) a droplet in air, (2) an air bubble in liqid, and (3) the two-fluid example of oil-in water where $\mu_2/\mu_1 = 1.2$ and $\rho_2/\rho_1=0.32$ were used. The abcissa is $A_1$ for cases (1) and (3), and $A_2$ for case (2). Shaded areas denote underdamped modes. The interval of $l$ values that contribute significantly to the force density in Eq.~(\ref{sigmaLeg}), $2\leq l \lesssim \alpha$, are indicated  (slanted and horizontal dashed lines) for air-water and oil-water, respectively, when $\lambda_0=1064$nm.}
  \label{fig_overdamp}
\end{figure}

For a droplet in air, $\mu_2\ll\mu_2$ and $\rho_2\ll \rho_1$, so we may safely replace air with vacuum. The criterium for underdamped oscillation for any mode whose index $l$ is then
\begin{align}
  A_1 >& \frac{(2 l^2-l-1)^2}{l(l^2+l-2)},  ~~~l\geq 2,
\end{align}
where
\[
  A_1=\frac{a\gamma}{\Delta \rho \nu_1^2},
\]
i.e., when
\begin{align}
  l<&\frac{1}{12}\Bigl\{A +2 \sqrt{36+24A+A^2} \notag \\
  &\times\cos\Bigl[\third \arctan\frac{6 \sqrt{3 A} \sqrt{648 + 441 A + 136 A^2 + 4 A^3}}{216 + 54 A + 36 A^2 + A^3}\Bigr]\Bigr\}  \notag \\
  \to & \left\{\begin{array}{cl}A/4, & \text{when }A\gg 1 \\ 1, & \text{when }A\ll 1\end{array}\right.\label{overdamp}
\end{align}
where the last line is asymptotes for large and small $A$.
When $A<25/8=3.125$, all modes $l>2$ are overdamped. This is illustrated in figure \ref{fig_overdamp}.

In the lingo of dimensionless numbers in fluid mechanics, 
\be
  A_1 = \frac{\mathrm{Ga}_1}{\mathrm{Eo}}
\ee
where $\mathrm{Ga}$, the Galilei number, is the ratio of gravitational to viscous forces, and $\mathrm{Eo}$, the E\"{o}tv\"{o}s number (virtually identical to the Bond number) is the ratio of bouyancy forces to capillary forces,
\[
 \mathrm{Ga}_1 = \frac{ga^3}{\nu_1^2}, ~~~\mathrm{Eo}= \frac{\Delta \rho a^2 g}{\gamma}.
\]

Bearing in mind the conclusion from \cite{brevik99,ellingsen12} that the number of significant terms in the $l$ sum is no more than order $\alpha$ (for the static case it can be significantly less), the behaviour of the droplet is sure to be essentially underdamped when $\alpha\lesssim A/4$, assuming $A\gg1$.

For the example of a water droplet in air ($\Delta\rho=997$kg/m$^3$, $\gamma=0.073$N/m, $\nu=1.01\times 10^{-6}$m$^2$/s), the value of $A$ is 
\[
  A_1^\text{water--air} = 72.5 \cdot a/1\,\mu\text{m}.
\]
For radii of a few microns, all important oscillations are thus underdamped. The condition that all significant modes be underdamped, $\alpha\lesssim A/4$ can be written
\[
  \lambda_0 \gtrsim \frac{8\pi n_2 \Delta \rho \nu^2}{\gamma}
\]
($\lambda_0$: laser wavelength in vacuum).
For water-air, this implies underdamped oscillations whenever
\[
  \lambda_0 \gtrsim 350\,\text{nm}
\]
which includes the full visible spectrum.

\subsubsection{Gas bubble in liquid}

In the opposite case, $\mu_2\gg\mu_2$ and $\rho_2\gg \rho_1$, the condition for mode $l$ to be underdamped is
\begin{align}\label{Abubble}
  A_2 >& \frac{(2l+1)^2(l+1)^4(l+2)}{l^5(l-1)},  ~~~l\geq 2,
\end{align}
where
\[
  A_2=\frac{a\gamma}{\Delta \rho \nu_2^2},
\]

As shown in Fig.~\ref{fig_overdamp}, the underdamped region makes up a smaller portion of the $l$-$A_2$ space than was the case in the $l$-$A_1$ space for the droplet. Clearly, viscosity has a somewhat greater influence on oscillations on the surface of bubble than on droplet.

One may ascertain from Eq.~(\ref{Abubble}) that all modes for a bubble are overdamped when
\[
  A_2<82.23.
\]
For an air bubble in water, $A_2=72.5 a/1\mu$m, hence it follows that there exists a range of very small radii for which all surface wave modes of an air bubble in water are overdamped, whereas those of a droplet of the same size are underdamped. For sufficiently large bubbles, however, $A>2025/8=253.125$, all relevant modes are underdamped whenever $\lambda_0 \gtrsim 350\,\text{nm}$, just as was found \emph{always} to be the case for droplets. Also this is indicated in figure \ref{fig_overdamp}.

\subsubsection{Two fluids}

Since droplet and bubble denote two limits of the general expression for the viscosity coefficient $\mu_l$, corresponding to $\mu_2/\mu_1, \rho_2/\rho_1\to 0$ and $\to \infty$, respectively, it is only to be expected that the critical damping curve of any two-fluid system where the properties of both fluids must be included will fall somewhere in between those of the limiting cases. Numerically this is found to be the case.

Due to the extremely low interfacial tension $\gamma$ for the oil-emulsion system, however, the nature of the critical damping curve for this set-up is found to be of no relevance. Whereas underdamped modes can only ever occur for values of $A_1$ in the order of $10^2$ or higher (see Fig.~\ref{fig_overdamp}), the numerical value found with the parameters used herein, chosen for ultra-low surface tension, is
\[
  A_1^\text{oil--water} \approx 0.00050 \cdot a/1\,\mu\text{m}
\]
($\Delta\rho=320$kg/m$^3$, $\gamma\approx 5\times 10^{-7}$N/m, $\mu_\text{oil}=1.2\mu_\text{water}$ -- see Refs.~\cite{ward06,ward07,mitani02}). The droplet radius required for $A_1\sim 10^2$ is thus far into the macroscopic regime. We can safely conclude that all surface wave modes of the oil emulsion are underdamped.

In conclusion, surface oscillations of water droplets in air are entirely underdamped, of air bubbles can be either under- or overdamped, and of oil-emulsions in water are entirely overdamped. This holds for all microscopic particles illuminated by laser light in the IR, visible or near-UV spectrum.
 
\subsection{Gaussian beam profile}

The electromagnetic fields in a Gaussian beam have no simple closed expression, but have been found to high order in expansion parameter $\lambda/w_0$ ($w_0$ is waist beam width) when this is small \cite{barton89b}. Taking the example of the linearly polarised fields given in Ref.~\cite{davis79} (we use the formalism of Ref.~\cite{barton88}) and adapting to circular polarisation ($E_y=iE_x$, etc.) leading order expressions are found to be
\begin{subequations}\label{Edef}
\begin{align}
  E_x^i \approx& E_0\psi_0^*e^{ikz}; ~~~ E_y^i = iE_x^i;\\ 
  E_z^i \approx& -\frac{2Q^*}{kw_0^2} (x+iy)E_x^i;\\
  \mathbf{H}^i \approx& -\frac{in_2}{\mu_0 c}\mathbf{E}^i
\end{align}
\end{subequations}
with
\be
  \psi_0 = iQ e^{-iQ(x^2+y^2)/w_0^2}; ~~~Q = \frac1{i+2z/kw_0^2}.
\ee
This is a solution to Maxwell's equations, modulo higher order corrections.
Superscript $i$ on $E$ and $H$ denotes incident fields. This now describes the incoming fields from the laser beam.

Note that $E_0$ is not the complex amplitude of the E-field in the above, but is chosen so that $\langle\mathcal{E}^2\rangle=E_0^2$ at $x=y=z=0$.

We define another dimensionless parameter $\kappa$, being the number of wavelengths per beam width
\be
  \kappa = kw_0.
\ee
In typical situations both $\alpha$ and $\kappa$ can far exceed unity. The field expressions of Eq.~\ref{Edef} are truncated at leading order in $1/\kappa$, hence valid for $\kappa\gg 1$. This ensures that field expressions remain good many wavelengths away from the beam focus. We will see in the following that the condition for the approximate field expressions to hold within the entire droplet or bubble is
\[
  \kappa^2 \gg \alpha,
\]
which is thus the criterion for the accuracy of the theory in the following. This is a restriction on the relative magnitudes of our three length-scales: radius, beam width and wavelength,
\[
  \frac{\lambda_0}{w_0} \ll 2\pi n_2 \frac{w_0}{a}.
\]

\subsection{Internal field and explicit force density}\label{sec_fields}

In spherical coordinates the internal fields inside the particle, at coordinate $r,\theta$, are given by \cite{barton89}
\begin{subequations}\label{generalFields}
\begin{align}
  E_r^w=& \frac{E_0}{\varrho^2} \sum_{l=1}^\infty\sum_{m=-l}^ll(l+1)\tc_{lm}\psi_l(\bnar)Y_{lm}(\Omega)\\
  E_\theta^w=& \frac{\alpha E_0}{\varrho}\sum_{l=1}^\infty\sum_{m=-l}^l \Bigl[\bn \tc_{lm}\psi'_l(\bnar)\partial_\theta Y_{lm}(\Omega)\notag \\&
  -\frac{\td_{lm}}{n_2}m\psi_l(\bnar)\frac{Y_{lm}(\Omega)}{\sin\theta}\Bigr]\\
  E_\phi^w=& \frac{i\alpha E_0}{\varrho} \sum_{l=1}^\infty\sum_{m=-l}^l \Bigl[m\bn \tc_{lm}\psi'_l(\bnar)\frac{Y_{lm}(\Omega)}{\sin\theta}  \notag \\  &
  -\frac{\td_{lm}}{n_2}\psi_l(\bnar)\partial_\theta Y_{lm}(\Omega)\Bigr]
\end{align}
\end{subequations}
where $\Omega=(\theta,\phi)$, $\varrho=r/a$, and the coefficients as defined by Barton et al.\ are \cite{barton89} (we will define more suitable versions below)
\begin{subequations}
\begin{align} 
  \tc_{lm} =& i\tA_{lm}[\bn^2\psi_l(\bna)\xi_l^{(1)\prime}(\alpha)-\bn\psi'_l(\bna)\xi_l^{(1)}]^{-1},\\
  \td_{lm} =& i\tB_{lm}[\psi_l(\bna)\xi_l^{(1)\prime}(\alpha)-\bn\psi'_l(\bna)\xi_l^{(1)}]^{-1}.
\end{align}
\end{subequations}
The incident field $\mathbf{E}^i$ is contained in the quantities
\be\label{AB}
  \begin{array}{c}\tA_{lm}\\ \tB_{lm}\end{array} = \frac1{l(l+1)\psi_l(\alpha)}\int\left.\begin{array}{c}E_r^i/E_0\\H_r^i/H_0\end{array}\right.Y_{lm}^*(\Omega)\rmd \Omega,
\ee
where incident fields are evaluated at $r=a^+$ and the integral is over all solid angles. Here $H_0=E_0/(c\mu_0)=c\varepsilon_0E_0$. 

For a circularly polarized Gaussian beam the radial incident field is (as always truncating to leading order in $\lambda/w_0\sim 1/\kappa$)
\begin{align}
  &\frac{E_r^i}{E_0} = \frac{E_x^i\sin\theta e^{i\phi} + E_z^i\cos\theta}{E_0}\notag\\
  &= \sin\theta e^{i\phi}\frac{E_x}{E_0}\Bigl[1-\frac{2i\alpha }{\kappa^2}\cos\theta+...\Bigr]\notag \\
  &= \frac{\kappa^4\sin\theta e^{i\phi}}{(\kappa^2+2i\alpha\cos\theta)^2}\exp\Bigl[\frac{i\alpha\kappa^2\cos\theta-\alpha^2(1+\cos^2\theta)}{\kappa^2+2i\alpha\cos\theta}\Bigr]
\end{align}
We see immediately that the criterion for the correction terms in the approximate expressions for the incoming fields to be small is that $\kappa^2\gg \alpha$ as stated above.

To calculate $\tA_{lm},\tB_{lm}$ 
using Eq.~\eqref{AB}
we use the relation
\[
  Y_{lm}(\Omega) = \sqrt{\frac{2l+1}{4\pi}\frac{(l-m)!}{(l+m)!}}P_l^m(\cos\theta)
  e^{im\phi}
\]
and since 
$E_r^i= e^{i\phi} f(\theta;\alpha,\kappa)$ 
only $m=1$ coefficients are nonzero:
\[
  \int_0^{2\pi}\rmd \phi e^{i\phi}e^{-im\phi} = 2\pi\delta_{m1}.
\]
We hence obtain the expressions
\bs
  \begin{align}
    \tA_{lm} =& \frac{\sqrt{(2l+1)\pi}\delta_{m1}}{[l(l+1)]^{3/2}\psi_l(\alpha)}\Phi_l(\alpha,\kappa); \\
    \tB_{lm} =& -in_2   \tA_{lm} ,
  \end{align}
\es
and
\begin{align}
    \Phi_l(\alpha,\kappa)=&\int_{-1}^1 \rmd u \frac{\kappa^4\sqrt{1-u^2}P_l^1(u)}{(\kappa^2+2i \alpha u)^2}\notag \\
      &\times\exp\Bigl[\frac{i\alpha\kappa^2u-\alpha^2(1+u^2)}{\kappa^2+2i \alpha u}\Bigr] ,
\end{align}
which must be evaluated numerically. In the limit where the laser waist tends to infinity we find
\be
  \lim_{\kappa\to\infty}\Phi_l(\alpha,\kappa) = \frac{2i^{l+1}}{\alpha^2}l(l+1)\psi_l(\alpha)
\ee
which gives back the known plane wave case~\cite{lai89,ellingsen12}.

With this we are ready to insert our expressions to obtain the force density of Eq.~\eqref{sigmaEM} and project onto the basis of Legendre polynomials using (\ref{sigmal}). With some straightforward but tedious algebra we obtain
\begin{widetext}
  \begin{align}
    \sigma_l=&\frac{\varepsilon_0n_2^2E_0^2(\bn^2-1)}{32}\sum_{m=1}^\infty\sum_{n=1}^\infty\Bigl\{ c_mc_n^*\psi_m(\bna)\psi_n  (\bna) I_{lmn} 
    +\alpha^2\Bigl[c_mc_n^*\psi_m'(\bna)\psi_n'(\bna)+d_md_n^*\psi_m(\bna)\psi_n(\bna)\Bigr]M_{lmn}\notag \\
    &+i\alpha^2\Bigl[c_md_n^*\psi_m'(\bna)\psi_n(\bna)-d_mc_n^*\psi_m(\bna)\psi_n'(\bna)\Bigr]N_{lmn}\Bigr\}.\label{bigsigma}
  \end{align}
  with the constant coefficients
  \begin{subequations}\label{IMN}
    \begin{align}
      I_{lmn}=&\frac{(2l+1)(2m+1)(2n+1)}{m(m+1)n(n+1)}\int_{-1}^1\rmd uP_l(u)P_m^1(u)P_n^1(u),\\
      M_{lmn}=&\frac{(2l+1)(2m+1)(2n+1)}{[m(m+1)n(n+1)]^2}\int_{-1}^1\rmd uP_l(u)\Bigl[(1-u^2)P_m^{1\prime}(u)P_n^{1\prime}(u)
      +\frac{P_m^1(u)P_n^1(u)}{1-u^2}\Bigr]\\
      N_{lmn}=&-\frac{(2l+1)(2m+1)(2n+1)}{[m(m+1)n(n+1)]^2}\int_{-1}^1\rmd uP_l'(u)P_m^1(u)P_n^1(u)
    \end{align}
  \end{subequations}
\end{widetext}
and where we define more handy $c$ and $d$ coefficients,
\begin{subequations}
  \begin{align}
    c_l =& \frac{\Phi_l(\alpha,\kappa)}{\psi_l(\alpha)}\frac1{\bn\psi_l(\bna)\xi_l^{(1)\prime}(\alpha)-\psi'_l(\bna)\xi_l^{(1)}}\\
    d_l =&   \frac{\Phi_l(\alpha,\kappa)}{\psi_l(\alpha)}\frac1{\psi_l(\bna)\xi_l^{(1)\prime}(\alpha)-\bn\psi'_l(\bna)\xi_l^{(1)}}.
  \end{align}
\end{subequations}
Exact expressions for $I_{lmn}, M_{lmn}$ and $N_{lmn}$ are given in appendix \ref{app_IMN}

Calculating $\sigma_l$ is the major numerical task, whereupon insertion into \eqref{hl} or \eqref{motion} quickly yields the static and transient surface deformations.

\subsection{Note on intensity and power}

Experimental values are normally not given for $E_0$, but rather for intensity or laser power. The intensity is given by the Poynting vector ($\langle\cdots\rangle$ denotes average over one optical period as usual),
\begin{align*}
  I =& \langle \mathcal{S}_z\rangle = \langle \bc{E}\times\bc{H} \rangle_z=\half\mathrm{Re}\{ \mathbf{E}\times\mathbf{H}^* \}_z=\frac{n_2|E_x|^2}{\mu_0 c}\\
  =&\frac{n_2|Q|^2E_0^2}{\mu_0 c}e^{-2|Q|^2(x^2+y^2)/w_0^2}\\
  \equiv& I_0|Q|^2e^{-2|Q|^2(x^2+y^2)/w_0^2}
\end{align*}
where we have made the observation that $\mathrm{Re}\{iQ\}=|Q|^2$.

The power is given by the integral of the Poynting vector across the beam cross section, and is independent of $z$ as it should:
\be
  P = 
  \half \pi w_0^2 I_0 =
  \half \pi\varepsilon_0 c n_2w_0^2E_0^2.
\ee
Equation \eqref{bigsigma} can then be written 
\begin{align}
  \sigma_l&=\frac{n_2I_0(\bn^2-1)}{32c}\sum_{m=1}^\infty\sum_{n=1}^\infty\{\cdots\}\notag \\
  &=\frac{n_2P(\bn^2-1)}{16\pi cw_0^2}\sum_{m=1}^\infty\sum_{n=1}^\infty\{\cdots\}.
\end{align}

The effective power is the intensity of the laser light integrated over the central cross section of the sphere,
\be
  P_\text{eff}
  =\frac{\pi I_0 w_0^2}{2}(1-e^{-2a^2/w_0^2}).
\ee
In the case $w_0\gg a$ we obtain the obvious limit
\be
  P_\text{eff}^{\kappa\to \infty} = \pi  a^2I_0.
\ee
When comparing the case of a narrow laser beam $\kappa < \alpha$ to one of infinite width, we will quote the effective power, which is the comparable unit between the two. 

\subsection{Necessary numerics}

The numerical calculation involved is thus twofold. The coefficients $I_{lmn},M_{lmn}$ and $N_{lmn}$ may be calculated once and for all and is numerically the heaviest part, and secondly the integrals $\Phi_l(\alpha,\kappa)$ must be calculated for a given set of parameters. 

The sums in Eq.~\eqref{bigsigma} can each be truncated after $\mathcal{O}(\alpha)$ terms, and similarly the number of $\sigma_l$s of significant magnitude. Explicit calculated values of the coefficients of Eq.~\eqref{IMN} (quoted in appendix) can be written as triple sums with a total of $lmn$ terms. We see that the problem initially has a terrible scaling with increasing $\alpha$; the calculational cost increases as $\alpha^6$. It is not untypical with values of $\alpha$ in the order of hundreds, and calculation time quickly becomes forbidding. 

A scaling closer to $\alpha^3$ can be obtained by approximating the integrals \eqref{IMN} using a trigonometric approximation for the legendre polynomials~\cite{thorne57}. Upon intergration this approximation is only accurate to about 20\%, however, so the search for a quicker algorithm continues, for example by taking the approximation to higher order.

\section{Numerical examples}\label{sec_numerics}

\begin{figure*}[tb]
  \includegraphics[width=.75\textwidth]{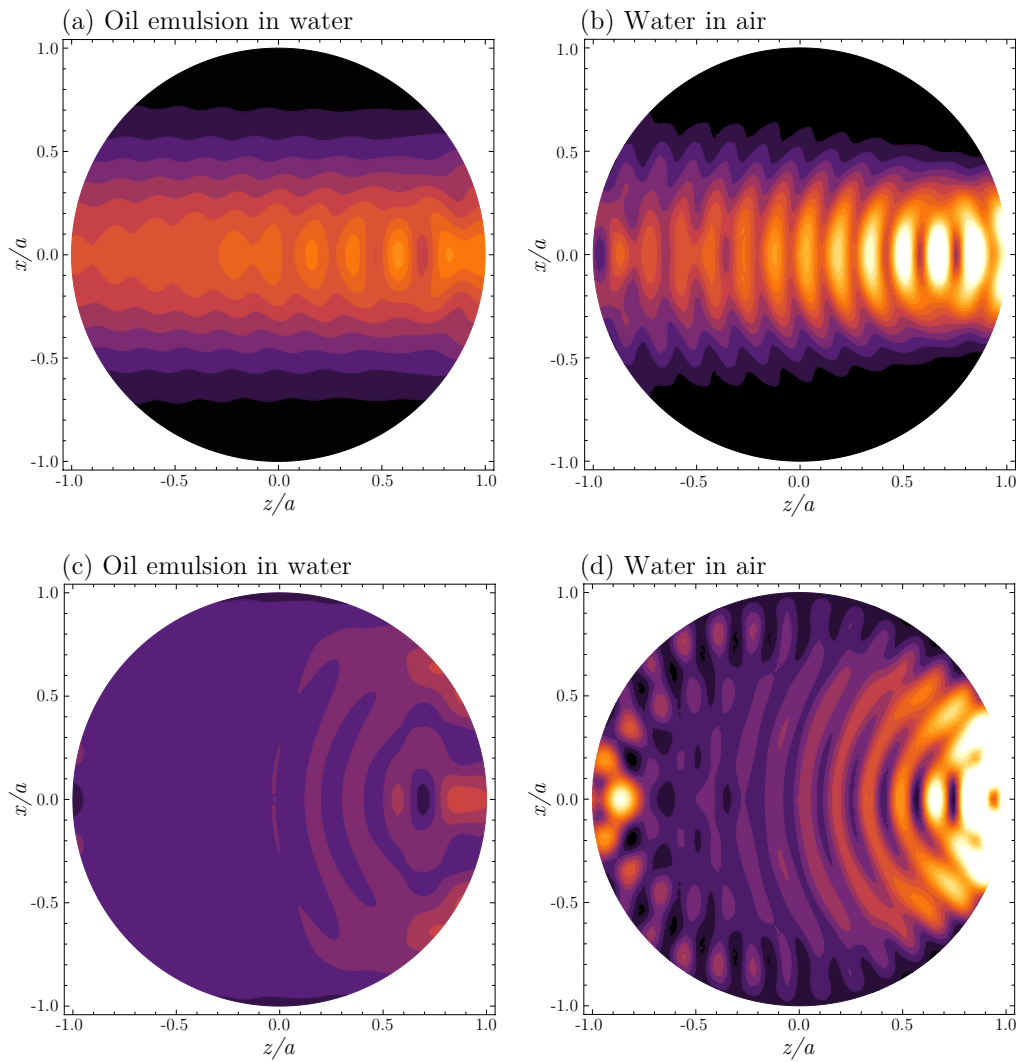}
  \caption{Absolute square of electric field $|\mathbf{E}^i|^2$ inside droplets with lower and higher dielectric contrast. The droplet radius in this example is $a=2\mu$m. Beam waist at focus in (a) and (b) is $1.5\mu$m ($\kappa = 11.8$), and infinite in (c) and (d). Vacuum wavelength is $\lambda_0=1064$nm, so $\alpha=15.7$ in all cases. Contours are at the same levels in (a) and (b) ($15$ contours between $0$ and $2.5E_0^2$; white areas: $E^2>2.5E_0^2$), and in (c) and (d) ($15$ contours between $0$ and $6E_0^2$; white areas: $E^2>6E_0^2$). Peak values are $1.43E_0^2$ (a), $4.35E_0^2$ (b), $2.07E_0^2$ (c), and $15.75E_0^2$ (d).}
  \label{fig_fields}
\end{figure*}

In the following we will provide a number of numerical examples of the above theory. We focus on three different physical situations; a water droplet in air \cite{zhang88}, an air bubble in water \cite{anand11}, and an oil emulsion droplet in water \cite{ward07}.

\subsection{Internal field intensity}

Maxwell's stress tensor is intimately connected to the electric field intensity, as is immediately clear from Eq~(\ref{sigmaEM}). The higher the intensity, that is, the mean square field, near the spherical boundary, the larger the local optical force density. It is of interest therefore to compare the electric field intensity within droplets of the two different dielectric contrasts found in the water-air system (higher contrast, $n_2=\bn=1.33$) and oil emulsion system (low contrast $n_1 = 1.41$, $\bn=1.06$). Results are shown in Fig.~\ref{fig_fields} for an intermediate size droplet ($\alpha=15.7$) chosen in respect to the wavelength so that scattering patterns be neither trivial nor unneccessarily complex.

From the prefactor $\bn^2-1$ in Eq.~\eqref{sigmaEM} it seems that the local force density on the droplet surface should be proportional to the dielectric contrast $\Delta n$, but in truth the scaling is stronger than this because of focussing effects from refraction. A droplet with higher dielectric contrast such as water in air will act as a lens, focussing the incoming light onto its rear interface, creating spots of significantly higher intensity than average. The difference between the two cases is clear in figure \ref{fig_fields}, both for a beam that is narrower than the droplet's diameter (Fig.~\ref{fig_fields} a and b), and for a wide beam (Fig.~\ref{fig_fields} c and d). In the figure we compare a situation of low dielectric contrast (oil emulsion in water, $n_1=1.33, n_2 = 1.41$) in a and c to the higher contrast system of water in air in b and d. 

Because the deforming force goes like light intensity at the dielectric boundary, the shapes obtained can be qualitatively different due to the different contrast in otherwise comparable circumstances. A corollary of this is that to correctly treat systems with even higher dielectric contrast in the static case where the droplet is still illuminated after it is deformed (as opposed to a short pulse), a more general scattering theory should be drawn upon for full accuracy. Such a theory does not exist to date to our knowledge, and will be the topic of future studies.

\subsection{Static deformations}

\begin{figure*}
  \includegraphics[width=.7\textwidth]{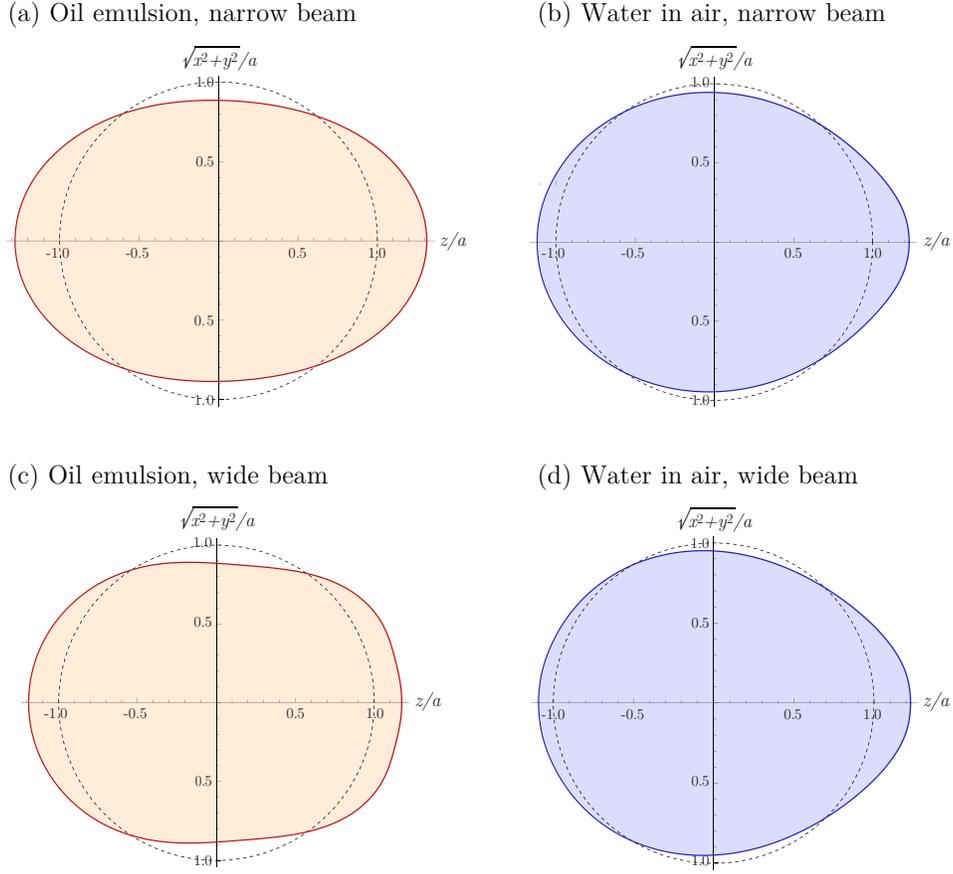}
  \caption{Approximate static shapes for the same configurations as in figure \ref{fig_fields}. The laser power in (a) is $7$\,mW ($P_\text{eff}=6.8$\,mW), and the effective power in (c) is $4$ times that of (a). The laser intensity in (d) is $6$\,W\,$\mu$m$^{-2}$ ($P_\text{eff}=75$W) and the effective power in (b) is $1/3$ that in (a) ($25$\,W). Droplet radius is $a=2\mu$m in all examples.}
  \label{fig_shapes}
\end{figure*}

The static deformations obtained for the four configurations in figure \ref{fig_fields}, are shown in the corresponding panels of figure \ref{fig_shapes}. The four cases illustrate well the different properties observed in changing beam width and for different optical contrast $\bn$. 
The static shapes should be considered semi-quantitative only, since the EM fields used are only zeroth order. For the oil-emulsion system, the optical contrast is so low that scattering is modest, and the approximation is not expected to be bad. For water-air there will clearly be corrections. A redeeming feature is, however, that the greatest deformation of the surface is found on the side of the droplet facing away from the laser beam, onto which the light is focussed, whereas the irradiated side is still close to hemispherical with a somewhat smaller radius of curvature. Hence we can expect the focal point of the light to be moved slightly towards the centre of the droplet, and the static deformation therefore to be slightly overestimated by our theory. An improvement of this particular aspect of the treatment is a goal for the future.

\subsubsection{Wide vs narrow beam, high vs low optical contrast}

We consider now the deformations that can be obtained with a narrow beam ($w_0=1.5\mu$m as before, and $a=2\mu$m) and a wide beam, applied to the low and high optical contrast systems, oil-emulsion and water droplets, respectively. 

For the oil-emulsion system we take values similar to those used in \cite{ward06}, a total laser power of $7$\,mW in the narrow beam case. Not all of the light impacts on the droplet, so the effective power is $6.8$\,mW. The corresponding deformation is shown in Fig.~\ref{fig_fields}a. Just like the squared $E$-field the deformation is approximately front-back symmetrical, and the resulting shape is reminiscent of a melon. As figure \ref{fig_fields}a shows, the laser field propagates through the oil droplet almost unchanged, and the resulting deformation bears witness to the intensity profile being approximately Gaussian both at front and rear. 

The oil emulsion does not focus the light much due to low optical contrast, so when the beam width is much wider than droplet dimensions, the light intensity ends up being near uniform thoughout the droplet, as seen in Fig.~\ref{fig_fields}c. A uniform light intensity cannot deform an incompressible droplet, only intensity differences can. It is necessary therefore to increase the effective power of the laser in order to obtain similar field strengths and hence similar deformation magnitude as for the narrow beam. When increasing the effective power by a factor $4$, the deformation that results is that of figure \ref{fig_shapes}c. The local intensity maxima seen in Fig.~\ref{fig_fields}c, on the symmetry axis at rear wall, and in a circular band approximately $40^\circ$ from this, show up as corresponding deformations in Fig.~\ref{fig_shapes}c, resulting in what might be termed a beetle shape. 

\begin{figure}[htb]
  \includegraphics[width=2.3in]{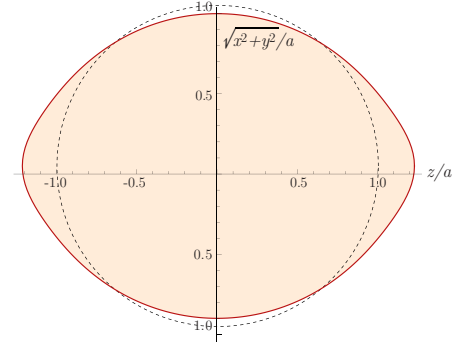}
  \caption{Approximate ``lemon'' shape of oil-emulsion with very narrow laser beam. We use $\lambda=385$nm as used in \cite{ward06,ward07}, radius $a=2.5\mu$m ($\alpha=54.3$) and waist $w_0=0.4\mu$m ($\kappa=8.7$) so $\kappa^2/\alpha=1.4$, thus somewhat beyond the scope of the theory.}
  \label{fig_lemon}
\end{figure}

It is prudent to note at this point that deformations of oil-emulsion systems are highly dependent on exact optical contrast, laser width, laser wavelength and radius. A slightly different refractive index difference would scatter the light differently in Fig.~\ref{fig_fields}c, say, resulting in a different shape in Fig.~\ref{fig_shapes}c. Likewise, an even narrower beam than that used herein (which, remember, can only be adequately described in our formalism so long as $\kappa^2\gg \alpha$), results in a lemon shape instead. We have approximated this in figure \ref{fig_lemon} -- although $\kappa^2/\alpha\sim 1$ in the figure, the shape is expected to be approximately right. In order to have a beam significantly narrower than $a$ without far exceeding the bounds of our theory, we choose compromise variables $\lambda_0 = 385$nm (used in \cite{ward06,ward07}), $a=2.5\mu$m, $w_0=0.4\mu$m with oil-emulsion parameters otherwise the same. A lemon shape results.

Since the surface tension of a water-air interface is much higher than that of the oil-emulsion system, correspondingly higher laser power is required for deformations to be comparable. We choose a light intensity that is similar to that of Refs.~\cite{zhang88} and \cite{ellingsen12}, $6$W$\mu\mathrm{m}^{-2}$, and which gives a clear deformation even for such a small droplet as $a=2\mu$m (much smaller than the droplets used by Zhang and Chang \cite{zhang88} where the lowest of the intensities used is $4$W$\mu\mathrm{m}^{-2}$). 
While for water droplets this is a very small size, it is the droplet size used in typical emulsion systems \cite{ward06, moller09}, so we use it for ease of comparison.
The wide beam is strongly focussed onto the rear droplet perimeter, resulting in an intensity enhancement by a factor $\approx 16$. Larger droplets (i.e., larger $\alpha$) can result in even larger enhancements -- Zhang and Chang report enhancement factors $>200$. 

The effects of the higher optical contrast in the water--air case are curious to note. Contrary to the low--contrast case, deformations are now \emph{larger} for the wide beam than the narrow, when effective powers are the same. To get the similar deformation amplitudes seen in Fig.~\ref{fig_shapes}b and d, effective laser power had to be further increased by a factor 3 for the narrow beam. The second curious observation is that the deformation is qualitatively almost identical for narrow and wide beam in this case. The reason may be gathered from the intensity plots in Fig.~\ref{fig_fields}b and d. Both have intensity maxima near the axis on the rear surface. The intensity enhancement is greater for the wide than the narrow beam (maximum $\approx 16E_0^2$ compared to $\approx 4E_0^2$ for this example), but for droplet shaping this is compensated by significant field intensity values along the whole droplet surface.

We note finally a feature that is to the advantage of our approximate theory where electric fields for a sphere are used in static (trapping) deformation calculations: the front (illuminated) face of the water droplets are distorted from the spherical shapes much less than the rear surface, so corrections due to slightly different light scattering patterns are much smaller than if the opposite were the case (this can happen e.g.\ with a Bessel beam under special circumstances, cf.\ \cite{ellingsen12}, but never with a Gaussian beam).

\subsubsection{Bubbles vs droplets}

\begin{figure*}[htb]
  \includegraphics[width=\textwidth]{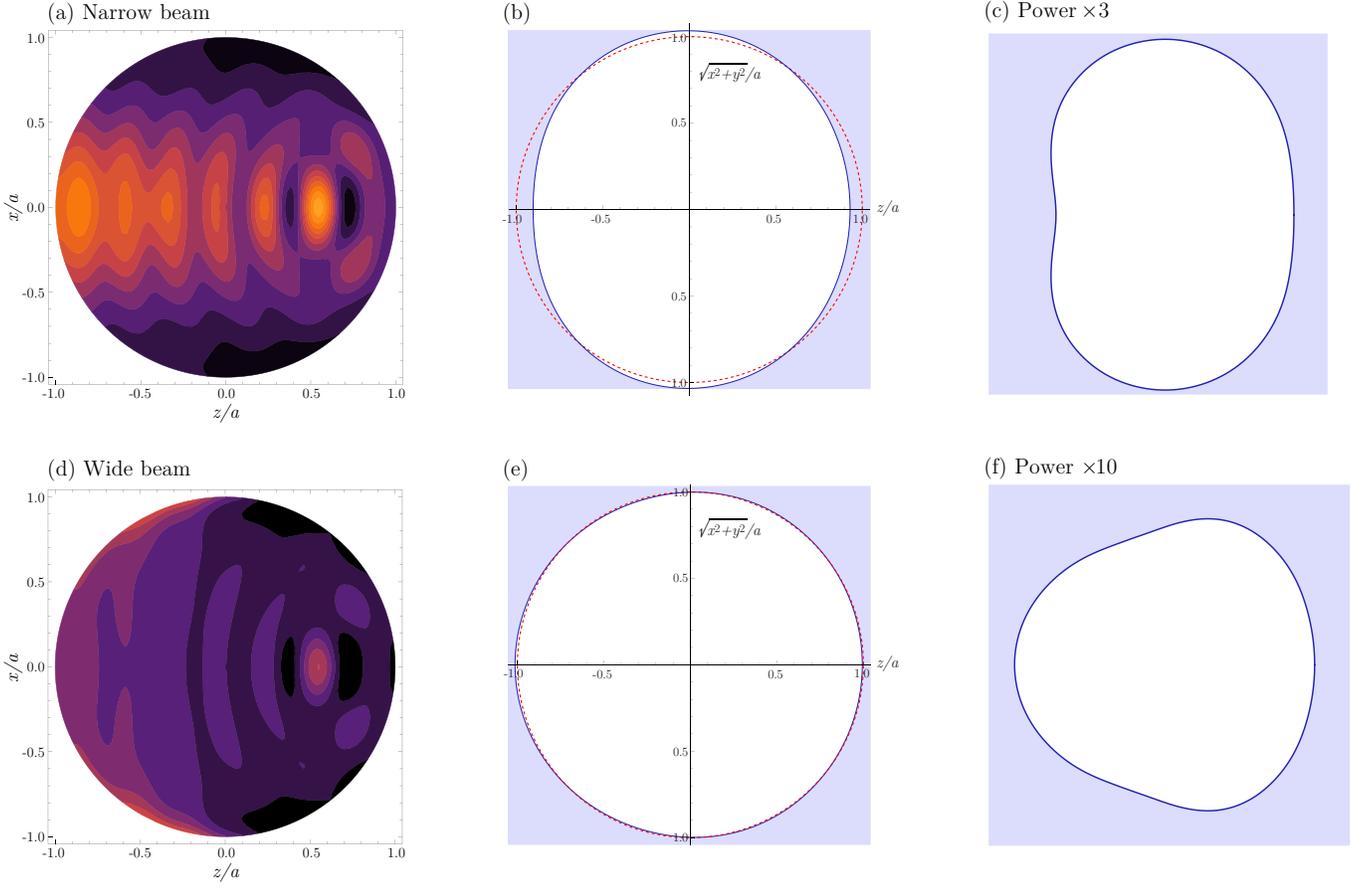}
  \caption{Bubble of air in water, for the same cases as considered in figures \ref{fig_fields} and \ref{fig_shapes}: a narrow beam (a-c) and a wide beam (d-f). Field intensity contour plots a and d are scaled the same as corresponding plots in figure \ref{fig_fields} for direct comparison. Figures b and e show deformations using the same laser powers as used for water droplets in figure \ref{fig_shapes} b and d, respectively. Panels c and f show the same deformations as panels b and e, respectively, but with power boosted by the indicated factor.}
  \label{fig_bubble}
\end{figure*}

The apparently minor operation of interchanging the inner and outer medium of illuminated body and exterior can give dramatically different results. We consider in this section the technologically interesting case of air bubbles in water, and compare it to our previous investigations of water droplets in air with the same laser specifications. Internal fields and deformations for cases which allow comparison with the droplet and emulsion sections are found in Fig.~\ref{fig_bubble}, to be discussed forthwith.

One immediate consequence of interchanging water and air is that the prefactor of Eq.~\eqref{bigsigma}, $\bn^2-1$, changes sign so that the surface force on the bubble interface is compressive. A simplistic prediction would be that the bubble would be squashed where the droplet bulged and vice versa.

The results in the two cases are more radically different than can be accounted for by this change of sign, however. The reason is found in the different light scattering properties of droplets and bubbles. While the droplet acted as a positive (converging) lens, focussing the light beam onto or near to its rear boundary, the bubble acts as a negative (diverging) lens, spreading the light out and away from its interior. This is strikingly visible upon comparing Fig.~\ref{fig_bubble}d with Fig.~\ref{fig_fields}d.

In figure \ref{fig_bubble} we study the case of an air bubble in water for comparison with the water droplet case in figures \ref{fig_fields} and \ref{fig_shapes}. Perhaps the most striking observation is that the bubble is deformed much less than the droplet under the same conditions, which is particularly noticeable for the wide beam. 

In order to visualise the kind of shapes which result with a narrow and broad beam, we also show the cases where the intensity is boosted sufficiently for deformations to be properly visible. Consistent with the observation that light is scattered away from the rear interface rather than focused onto it, the bubble is more clearly deformed on its front surface than its rear, in contradistinction to the case for water droplets. The shapes are qualitatively different from the ones previously observed, reminiscent of a sea urchin and an acorn, respectively.

A final note about optoacoustics is warranted. When considering a laser trapping a bubble in a surrounding liquid, the incoming and scattered laser beam will cause the liquid to contract slightly in regions of higher laser intensity. As the laser is switched on, a near-cylindrical sound wave is generated originating from the beam, propagating outwards. A detailed discussion may be found in Ref.~\cite{ellingsen11}. We have previously ignored the effect of electrostrictive compression, because it will be countered by an increase in mechanical pressure on an acoustical timescale, and plays no role in the dynamics after a time which we considered to be negligibly short. With a surrounding fluid, a soundwave originating from the laser beam can be reflected off nearby surfaces and arrive back at the bubble system some time after its emission. We have still neglected this effect since the sound wave's energy is distributed over a cylindrical surface, so a wave that has travelled far (and thus arrives late) will have comparatively little energy. In an unbounded fluid regime, it is of no consequence for later time, since the sound wave propagates away from the system studied.

\subsubsection{Comparison with experiment by M\o ller and Oddershede}

We are in a position now to compare the deformations calculated here with those measured by M\o ller and Oddershede \cite{moller09}. In their experiment the change of maximal radius of the droplet in the cross-beam plane, $r_\perp$, with increasing laser power. By using a simple theory based on apparently quite severe simplifications, they are able to predict the relation
\[
  \Delta r_\perp \approx -\frac{(F_\text{front}+F_\text{back})}{12\pi\gamma} = -A P
\]
where $F_\text{front,back}$ is the force on the front (rear) half of the droplet. In the system used in \cite{moller09} the constant $A$ is calculated to be $3.33\mu$mW$^{-1}$, and is found to be in surprisingly good agreement with the theory, especially since a simplified expression for the optical force is used, with the assumption that $w_0\ll a$, which is reasonably well satisfied in their system (they have $w_0\sim 1\mu$m and $a=2.3\mu$m to $2.9\mu$m). The system in Ref.~\cite{moller09} thus most resembles that shown in Fig.~\ref{fig_lemon}, producing a ``lemon'' shape.

Let us briefly compare the linear constant with the one calculated and measured by M\o ller and Oddershede. We have performed the calculation of droplet shapes using the data from \cite{moller09} ($\gamma=1.9\mu$N/m, $n_1=1.366, n_2 = 1.330$). We find that the shapes obtained, and the constant $A$ in particular, depends quite sensitively on the width of the laser waist, a parameter not given precicely in Ref.~\cite{moller09}, although $w_0<1\mu$m is indicated. For laser waists greater than droplet radius ($P$ interpreted as $P_\text{eff}$ in that case), we find $A\sim 2.3\mu$m/W, whereas numbers as high as $A\sim 4.5\mu$m/W are found when the beam waist is made so small that $\kappa^2/\alpha\sim 1$. As a general note, somewhat higher $A$ than the experimental one are found for narrow beams producing lemon shapes, whereas better agreement is found for beams producing ellipsiod shapes. 

With its narrow-waist laser, the geometries of Ref.~\cite{moller09} lie on or outside the boundary of validity for our theory, which is that $\kappa^2\gg \alpha$, hence perfect correspondence should not be expected. Their observed value lies in the range of values predicted by our theory. In effect, the approximate field expressions of Eqs.~\eqref{Edef} are valid when the beam does not broaden perceptibly compared to its waist within the confines of the droplet. It is likely that this does not hold for the experiment of M\o ller and Oddershede.

The present theory, however, would seem to indicate that the good correspondence between theory and experiment found in \cite{moller09} might be partly down to luck. The shape assumed by those authors is an ellipsoid, whereas the shape produced by a very narrow beam when the optical contrast is low we find rather to resemble the lemon shape of figure \ref{fig_lemon}. Moreover, three different droplet radii are investigated, and our results would seem to indicate that slightly different values of the slope $A$ are to be expected for different values of $\kappa^2/\alpha$. Ref.~\cite{moller09} does not give a quantitative bounds on the measured slope $A$ at a given confidence, so quantitative comparison with their experiment is difficult.

\subsection{Transient deformation: trapping beam vs short pulse}

\begin{figure*}[htb]
  \includegraphics[width=\textwidth]{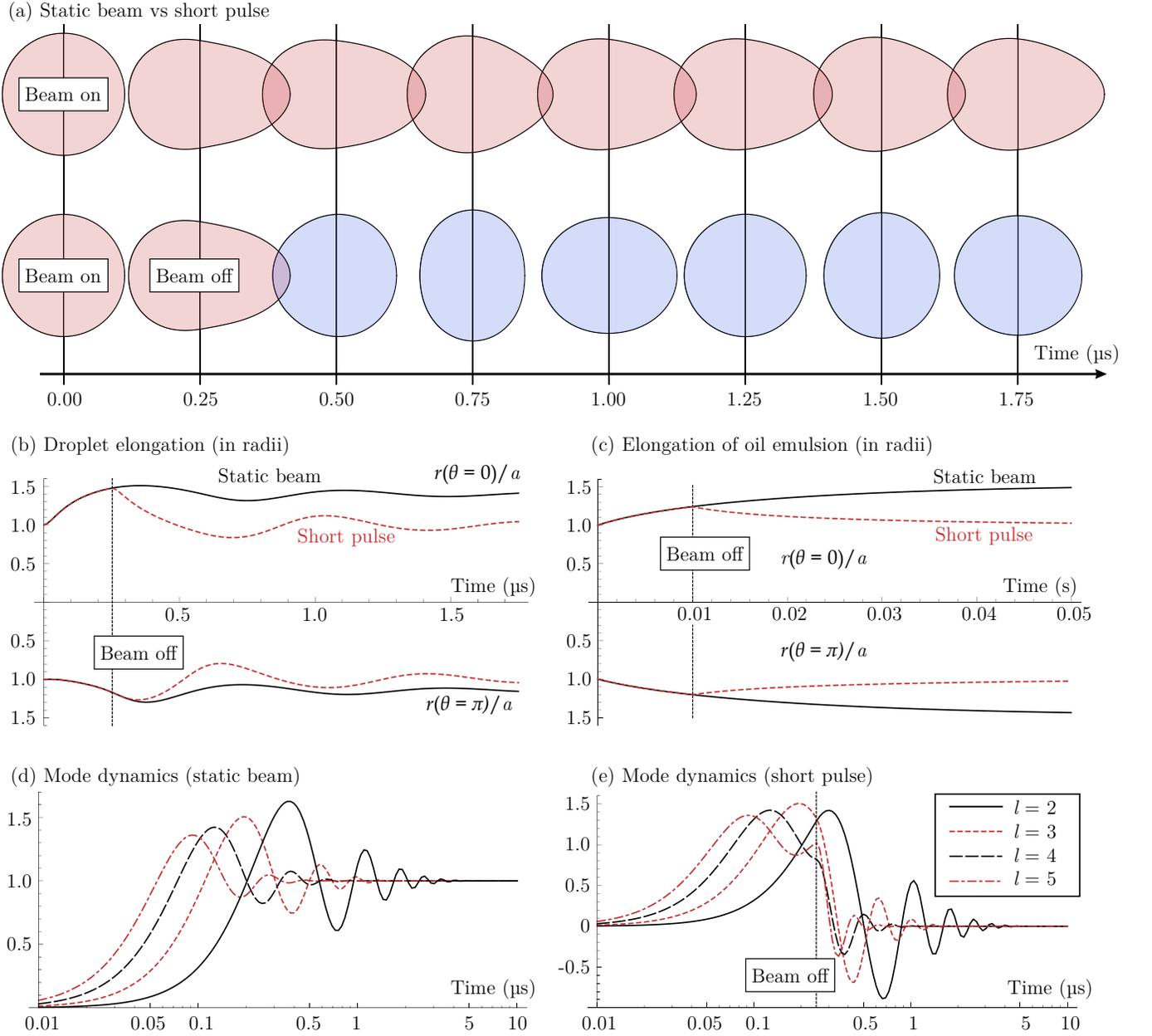}
  \caption{Dynamics of deformation for a water droplet, comparing the cases of static beam and a short pulse. (a) Water droplet in wide beam, as in Fig.~\ref{fig_shapes}d, but with slightly higher light intensity, $I_0=10$W$\mu$m$^{-2}$. Static beam (upper row) compared to beam of duration $0.25\mu$s (lower row). (b) Time development of droplet elongation for the above shapes, illustrated by $r(\theta=0,t)/a$ (rear of droplet, plotted above abcissa) and $r(\theta=\pi,t)/a$ (front of droplet where light enters, plotted below abcissa). Static beam turned on at $ty=0$ (solid) and pulse of duration $t_0=0.25\mu$s (dashed) are shown. Panel (c) shows the same as (b) but for the overdamped case of an oil emulsion in a narrow beam, same set-up as Fig.~\ref{fig_shapes}a, but with power boosted to $12$mW, for static beam switched on at $t=0$ (solid) and a pulse switched off again at $t_0=10$ms (dashed). Panel (d) and (e) show the time development of the four lowest deformation modes, as given in Eq.~(\ref{motion2}), for a static beam switched on at $t=0$ (d) and a pulse switched off again at $t_0=0.25\mu$s. The legend in panel (e) is for both panels (d) and (e).}
  \label{fig_dynamics}
\end{figure*}

We finally analyse the dynamics of droplet deformations in laser beams. We wish to study and compare the dynamics of  two different set-ups analysed in previous sections, a water droplet in air and the oil-emulsion in water. Dynamics of a gas bubble in water is found to not be very clearly and qualitatively distinguishable from that of a water droplet, and will not be given specific attention in this context. 

For a laser pulse abruptly turned on at $t=0$ and off again at $t=t_0$, Eqs.~(\ref{motion}) generalize to \cite{ellingsen12}
\bs\label{motion2}
  \begin{align}
   \frac{h_{l}(t)}{h_l(\infty)} =& 1-\Bigl(\frac{\mu_l}{\gamma_l}\sin\gamma_lt+\cos\gamma_lt\Bigr)\rme^{-\mu_l t}\notag \\
   &+\Theta(t-t_0)\Bigl\{1-\Bigl[\frac{\mu_l}{\gamma_l}\sin\gamma_l(t-t_0)\notag \\
   &+\cos\gamma_l(t-t_0)\Bigr]\rme^{-\mu_l( t-t_0)}\Bigr\}, ~~ \omega_l>\mu_l \label{motion2ud}
  \end{align}
  and
  \begin{align}
    \frac{h_{l}(t)}{h_l(\infty)} =& 1-\Bigl(\frac{\mu_l}{\Gamma_l}\sinh\Gamma_lt+\cosh\Gamma_lt\Bigr)\rme^{-\mu_l t}\notag \\
    &+\Theta(t-t_0)\Bigl\{1-\Bigl[\frac{\mu_l}{\Gamma_l}\sinh\Gamma_l(t-t_0)\notag \\
    &+\cosh\Gamma_l(t-t_0)\Bigr]\rme^{-\mu_l (t-t_0)}\Bigr\}, ~~ \omega_l<\mu_l 
  \end{align}
\es
for $t\geq 0$. 

An illustration of the dynamics of shapes is shown in Fig.~\ref{fig_dynamics}a, where a series of shapes of a water droplet is shown. The parameters are as in Fig.~\ref{fig_shapes}a, but with somewhat higher intensity, $I_0=10$W$\mu$m$^{-2}$, in order that the shapes be more clearly visible. In the top row is seen the case of a static beam switched on at $t=0$, and in the below row it is switched off again at $t_0=250$ns, whereupon the droplet returns to spherical shape when oscillations have died out, on a time scale of a few microseconds. 

We found in Section \ref{sec_damping} that the dynamics of water droplet and oil-emulsion are very different. The motion of the former is in practice always underdamped, the latter overdamped. Moreover, the time scale of a given mode $l$ is very different for the two. For an underdamped mode, a relevant timescale is its lifetime 
\be\label{lifetime}
  \tau_l^\text{droplet} = \frac{1}{\mu_l^\text{droplet}} = \frac{4.0\mu\mathrm{s}}{(2l^2-l-1)}
\ee
for water. The time scale of the overdamped oil emulsion, however, follows from Eq.~(\ref{motion}) wherein $\mu_l^2\gg\omega_l^2$, and
\begin{align}
  \tau_l^\text{emulsion} =& (\mu_l-\Gamma_l)^{-1}\approx \frac{2\mu_l}{\omega_l^2} = \frac{4\Delta \rho a \bar{\nu}}{\gamma}\frac{1}{lf(l)}\notag \\
  =& \frac{0.012\mathrm{s}}{lf(l)}
\end{align}
where $\bar{\nu}$ is the mean kinematic viscosity and $f(l)$ is a function of order unity (exact details can be derived from Eq.~(\ref{motion}) if of interest).

A striking difference between the water droplet and the oil emulsion is the timescale involved for the fluid dynamics. The motion of the water droplet reaches equilibrium within a couple of microseconds, whereas the oil emulsion takes up to a tenth of a second to find its final form. This difference is illustrated in figure \ref{fig_dynamics}b and c where the motion of the front and rear tips of the droplet are plotted. Above the abcissae the rear point position $r(\theta=0)/a$ (where the light exits) and below the front $r(\theta=\pi)/a$ (where the light enters). 

As noted also in Ref.~\cite{lai89, brevik99, ellingsen12}, the higher $l$ oscillation modes are short-lived compared to the lowest order modes, as also Eq.~(\ref{lifetime}) shows clearly. In Fig.~\ref{fig_dynamics}d and e we show the time evolution of the four lowest oscillation modes, as given by the right hand side of Eq.~(\ref{motion2ud}). By the time the $l=2$ mode reaches its maximum, only $l=3$ still contributes significantly to oscillations. Indeed, for the majority of the oscillation period of about $3\mu$s in the static beam case, only the $l=2$ mode contributes significantly to the time variation. That is not to say, of course, that the final shape of the droplet can be adequately described by the $l=2$ shape perturbation alone. 

Note that for a pulse of $250$ nanoseconds we may ignore compressibility effects due to electrostriction (see Ref.~\cite{ellingsen12b}). This becomes of importance when pulse duration is in the order of traversing time of a sound wave across the diameter, $t_s = 2a/c_s$ where $c_s$ is speed of sound. For liquid and gas bubble respectively, this time scale is approximately $4$ and $12$ns, which is much shorter than our pulse.

\section{Concluding remarks}

We have laid out a comprehensive theory for linear deformations of fluid micro particles in a laser beam of Gaussian profile. Three types of fluid systems were considered and compared: a water microdroplet in air, an air microbubble in water, and an oil-emulsion in water. The latter system has found several experimental applications due to its extremely low interface tension coefficient.

The fluid dynamics of the three types of fluid systems are surprisingly different. We show that dynamics of a water microdroplet in air is always underdamped, whereas the oil-emulsion always shows overdamped dynamics since the surface tension, which acts as a ``spring constant'' in the harmonic oscillator-type equations of motion, is very weak compared to the friction term from viscosity. A microscopic bubble in water can have significant contributions from both underdamped and overdamped modes, depending on radius and laser wavelength. 

Although the water droplet differs from the bubble in water only by an interchange of materials in the inner and outer region, their deformations from radiation pressure upon laser illumination are strikingly different. All else being equal, a laser which significantly disfigures a water droplet will typically hardly change a bubble at all. The reason being that a droplet acts as a positive lens, focussing the incoming light onto its shadow face, whereas a droplet is a negative lens, scattering the light away from its interior. A much higher laser power is therefore required to visibly deform a bubble than a droplet of the same size. 

We furthermore analyse the kind of shapes obtained when particles of the three different systems are illuminated by a wide laser beam (beam width $\gg$ radius) and fairly narrow beam (beam width $3/4$ of radius), respectively (the present theory is unable to handle beams which are much narrower than the droplet). Qualitatively different shapes are found, as we show in Figs.~\ref{fig_shapes} and \ref{fig_bubble}, as may be summed up as in Table \ref{tab_shapes}. 

\begin{table}[h!]
  \begin{tabular}{c|c c c}
    \hline
    &Droplet & Oil-emulsion & Bubble \\
    \hline  
    Wide beam & Egg & Beetle & Acorn\\
    Narrow beam & Egg & Melon/lemon & Sea urchin\\
    \hline
  \end{tabular}
  \caption{Qualitative shapes of fluid particles under illumination by narrow and wide laser beams. The shapes refer to those seen in Figs.~\ref{fig_shapes}, \ref{fig_lemon} and \ref{fig_bubble}.}
  \label{tab_shapes}
\end{table}

The light scattering pattern within the droplets and bubbles has everywhere been calculated as though the particle were spherical. This is near exact for laser pulses which are so short that the fluid mechanical motion happens after the light is off, but it is obvious that corrections must result in the case of a static beam, and the shapes calculated in this case must be considered semi-quantitative only. We expect this approximation to be fairly good in the case of oil-emulsion systems, where the dielectric contrast is small and focussing only very slight, as seen in Fig.~\ref{fig_fields} a and c. For water-air systems, however, the light intensity pattern is rather sensitive to exact shape. A redeeming feature is that the deformation of a water droplet is most prominent on the rear (shadow) side, and less on the side of the droplet impacted upon by the light. From the egg-like shapes in Fig.~\ref{fig_fields} b and d (which are exaggerated for illustration purposes) the correction would be due to the somewhat smaller radius of curvature on the side of the droplet facing the light, causing the focal point to move away from the rear wall somewhat. The expected effect might be a somewhat diminished deformation, yet qualitatively similar. We intend to return to this question more carefuilly in the near future.

The dynamics of a water droplet and an oil-emulsion were finally analysed. While not specific to a Gaussian beam profile, such a comparison of dynamics does not exist in the literature to our knowledge, and is of obvious interest to laser manipulation of different kind of fluid systems such as those considered. Because of the very low interfacial tension of the oil-emulsion droplet, its dynamics are slow compared to the water droplet. For the example of a droplet $4\mu$m across, the relevant timescales are tens of milliseconds and a few microseconds, respectively, i.e., they differ by $4$ orders of magnitude. We analyse more closely the dynamics of the water droplet, being underdamped. The longest wavelength oscillation mode $l=2$ was found to be the most long lived. Indeed, for approximately the latter $2/3$ of its lifetime, this mode is the only one contributing significantly to dynamics. The static shapes of the droplets, however, can have significant contributions from modes all the way up to $l=\alpha$ ($\alpha$ is the number of wavelengths per circumference), although surface tension tends to smear out the short-wavelength perturbations and lending emphasis to lower $l$ modes.

\subsection*{Acknowledgements}

This work was inspired and aided by discussions with Dr.\ Andy Ward, who has also provided data for the oil-emulsion system. We have furthermore benefited from discussions with Dr.\ Suman Anand and Dr.\ Kristian Etienne Einarsrud, and, as always, much help and encouragement was gained from interactions with Professor Iver Brevik. We gratefully acknowledge financial support from the Department of Energy and Process Engineering, Norwegian University of Science and Technology.

\appendix

\section{Exact values of $I,M,N$ coefficients}\label{app_IMN}

The coefficients in equation \eqref{IMN} can be evaluated explicitly. Let $\mathcal{I}_{lmn},\mathcal{M}_{lmn}$ and $\mathcal{N}_{lmn}$ be the integrals in equation \eqref{IMN}. We find the expressions
\begin{align}
  (\mathcal{I},\mathcal{M},\mathcal{N})_{lmn}=& 
 \underset{l'm'n'}{\hat{\mathbb{S}}}
  (i,m,n)_{lmn}^{l'm'n'}
\end{align}
where we use shorthand
\begin{align}
 \underset{l'm'n'}{\hat{\mathbb{S}}}=
 & \sum_{l'=0}^l\sum_{m'=1}^m\sum_{n'=1}^n\binom{l}{l'}\binom{m}{m'}\binom{n}{n'}\notag \\
 &\times\binom{\frac{l+l'-1}{2}}{l}\binom{\frac{m+m'-1}{2}}{m}\binom{\frac{n+n'-1}{2}}{n}\notag \\
 &\times e(l+l')e(m+m')e(n+n'),\\
  \Lambda=& l+m+n, ~~\Lambda'= l'+m'+n',
\end{align}
and use the even/odd selectors
\be
  e(n)=\begin{cases}1, &n \text{ is even}\\0, &n \text{ is odd} \end{cases},~~ 
  o(n)=\begin{cases}0, &n \text{ is even}\\1, &n \text{ is odd} \end{cases}.
\ee
The specific summands are
\bs
  \begin{align}
    i_{lmn}^{l'm'n'}=&2^{\Lambda+2}e(\Lambda)(\Lambda^{\prime 2}-1)^{-1},\\
    m_{lmn}^{l'm'n'}=&2^{\Lambda+1}e(\Lambda)m'n'\Bigl[\frac{m'n'}{\Lambda'+1}-\frac{2m'n'-m'-n'-1}{\Lambda'-1}\notag\\
    &+\frac{(m'-1)(n'-1)}{\Lambda'-3}\Bigr],\\
    n_{lmn}^{l'm'n'}=&-2^{\Lambda}o(\Lambda)l'm'n'(\Lambda^{\prime 2}-2\Lambda')^{-1}.
  \end{align}
\es
The even/odd prefactors removes half the terms. Another observation is that all terms are zero for which one out of $l',m',n'$ is greater than the sum of the other two.

Explicit evaluation of these sums is numerically reasonable for moderate values of $l,m,n$, yet since the number of terms is of order $lmn$, and and since the sum in Eq.~\eqref{bigsigma} requires terms up to $l\sim\alpha$ and $m,n\sim l$, the full evaluation of $\sigma_l$ requires a number of terms which scales as $\alpha^6$. This quickly becomes forbidding for large radius-to-wavelength ratios. This is the geometrical optics limit, however, where other, approximate methods may be employed.

\section{Viscous coefficient}\label{app_Visc}

Following the recipe of Brevik and Kluge \cite{brevik99} we derive the approximate expression for the viscous damping coefficient, Eq.~\ref{mul}. Since the velocities are small, we can make use of a velocity potential as is standard in linear gravity wave theory, $\mathbf{v}=\nabla\varphi$. Incompressibility dictates $\nabla^2\phi=0$, and hence the internal and external solutions can be expanded in modes. Then
\begin{align*}
  \varphi_\text{int} =& \sum_{l=0}^\infty C_{l}P_l^m(\cos\theta)\rme^{im\phi}\varrho^l,\\
  \varphi_\text{ext} =& \sum_{l=0}^\infty D_{l}P_l^m(\cos\theta)\rme^{im\phi}\varrho^{-l-1}.
\end{align*}
The interface is characterized by the equation $S = a+h(\theta)-r=0$. The kinematic boundary condition at the interface is then given by \cite{wehausen60}
\[
  \frac{\mathrm{D}S}{\mathrm{D}t}=\dot{S}+(\mathbf{v}\cdot\nabla)S=0
\]
which, when keeping only terms of leading order in the small amplitude $h$, gives simply
\be
  \dot{\varphi}|_{r=a} = v_r |_{r=a}
\ee
wherewith we match the inner and outer solutions to find
\[
  D_l = -\frac{l}{l+1}C_l.
\]

Following Ref.~\cite{brevik99} we work in analogy to a damped harmonic oscillator. The time average of complete mechanical energy is given as $\overline{E}_\text{mech} = \int \rmd^3x  \rho \overline{v^2} $ (\cite{landau59} \S25), that is,
\[
  \overline{E}_\text{mech}
  = \rho_1\int_\text{int}\rmd^3x \overline{(\nabla\varphi_\text{int})^2}+\rho_2\int_\text{ext}\rmd^3x \overline{(\nabla\varphi_\text{ext})^2}.
\]
Consider in detail the internal part:
\begin{align*}
  \overline{E}_\text{mech}^\text{int}=&\pi\rho_1a \sum_{l=0}^\infty\sum_{l'=0}^\infty \sum_{m=-l}^{l}\sum_{m'=-l'}^{l'}\overline{C_{lm}C_{l'm'}^*}\\
  &\times\int_0^1\rmd \varrho\varrho^{l+l'}\int_0^{2\pi}\rmd\phi\rme^{i(m-m')\phi}\\
  &\times\int_{-1}^1\rmd x\Bigl\{\Bigl(ll'+\frac{mm'}{1-x^2}\Bigr)P_l^m(x)P_{l'}^{m'}(x)\\
  &+(1-x^2)[\partial_xP_l^{m}(x)][\partial_xP_{l'}^{m'}(x)]\Bigr\}.
\end{align*}
Corrections to the integrals due to surface deformations (upper limit of $\varrho$ integral is generally $1+h/a$) give rise to second order terms in the differential equation and are neglected.
Using the integral relations quoted and derived in appendix \ref{app_ints} all non-diagonal terms are found to be zero, and the result is a simple sum over $l$ and $m$ only:
\[
  \overline{E}_\text{mech}^\text{int}= 4\pi\rho_1 a\sum_{l=0}^\infty\sum_{m=-l}^{l}\frac{l\overline{|C_{lm}|^2}}{2l+1}\frac{(l+m)!}{(l-m)!}.
\]
A completely analogous calculation gives the external mechanical energy, and the sum is
\be
  \overline{E}_\text{mech} = 4\pi a \sum_{l=0}^\infty\sum_{m=-l}^{l} \frac{l\overline{|C_{lm}|^2}}{2l+1}\frac{(l+m)!}{(l-m)!}\Bigl(\rho_1+\frac{l\rho_2}{l+1}\Bigr).\label{Emech}
\ee
There is thus a mechanical energy uniquely associated with coefficient $C_{lm}$, hence with mode $l,m$. 

The time average of the dissipated energy within a volume may be expressed in terms of surface integrals bounding said volume, $\overline{\dot{E}}_\text{mech} = -\oint \rmd^2x  \mu \overline{\nabla v^2}\cdot \hat{\mathbf{n}}$  (\cite{landau59} \S16) where $ \hat{\mathbf{n}} $ is a unit vector normal to and pointing out of the bounding surface. There is no contribution from the surface at infinity, so we obtain
\begin{align*}
  \overline{\dot{E}}_\text{mech} 
  =& -\mu_1a^2\int \rmd\Omega \overline{\partial_rv_\text{int}^2}|_{\varrho=1}+\mu_2a^2\int \rmd\Omega \overline{\partial_rv_\text{ext}^2}|_{\varrho=1}\\
  =& -\frac{8\pi}{a}\overline{|C_{lm}|^2}\sum_{l=0}^\infty\sum_{m=-l}^{l} l\frac{(l+m)!}{(l-m)!}\Bigl[\mu_1l(l-1)\\  
  &+\mu_2(l+1)(l+2)\Bigr].
\end{align*}
Here $\int\rmd\Omega$ is an integral over all solid angles. Noting the relation (\cite{landau59} \S25)
\be
  \mu_{lm} = \frac1{2}\left|\frac{\overline{\dot{E}}_{\text{mech},lm}}{\overline{E}_{\text{mech},lm}}\right|,
\ee
equation (\ref{mul}) results. As for the special case considered by Brevik and Kluge \cite{brevik99}, the damping coefficient does not depend on $m$ (only modes $m=1$ are excited by the Gaussian, circularly polarized laser anyway, but the general result will be useful for future, more general work. For the linearly polarised laser considered in \cite{brevik99}, modes $m=\pm 1$ are excited, whereas an order $m_0$ Bessel beam excites modes $m=m_0$ \cite{ellingsen12}).

\section{Some integral relations}\label{app_ints}

To derive Eq.~(\ref{Emech}) we require the well known relations
\begin{align*}
  \int_0^{2\pi}\rmd\phi\,\rme^{i(m-m')\phi} =& 2\pi\delta_{mm'}\\
  \int_{-1}^1\rmd x\, P_l^m(x)P_{l'}^{m}(x) =& \frac{2}{2l+1}\frac{(l+m)!}{(l-m)!}\delta_{ll'}.\label{PPint}
\end{align*}
In addition we require the integral
\begin{align}
  I_{ll'}^m \equiv&  \int_{-1}^1\rmd x\, f(x)\equiv \int_{-1}^1\rmd x\,\Bigl\{\frac{m^2}{1-x^2} P_l^m(x)P_{l'}^{m}(x)\notag \\
  &+(1-x^2)[\partial_xP_l^{m}(x)][\partial_xP_{l'}^{m}(x)]\Bigr\} .
\end{align}
Using Legendre's equation
\[
 \Bigl[(1-x^2)\partial_x^2 - 2x\partial_x + l(l+1)-\frac{m^2}{1-x} \Bigr]P_l^m(x)=0
\]
and
\[
  (\partial_xP_l^{m})(\partial_xP_{l'}^{m})=\partial_x^2(P_l^mP_{l'}^{m}) -P_l^m(\partial_x^2P_{l'}^{m})-P_{l'}^m(\partial_x^2P_l^{m}),
\]
we may write
\[
  f(x)=\frac{1}{2}\{\partial_x(1-x^2)\partial_x P_l^mP_{l'}^{m} + [l(l+1)+l'(l'+1)]P_l^mP_{l'}^{m}\}.
\]
The integral of the first term from $-1$ to $1$ is clearly zero from the fundamental theorem, and so, 
inserting into Eq.~(\ref{PPint}) 
we find
\be
  I_{ll'}^m = \frac{2l(l+1)}{2l+1}\frac{(l+m)!}{(l-m)!}\delta_{ll'}.
\ee
This integral was worked out by R.~Kluge and used, but not quoted, in Ref.~\cite{brevik99} (I.~Brevik, private communication).

\end{document}